# Stoichiometry-induced ferromagnetism in altermagnetic candidate MnTe

Michael Chilcote[1], Alessandro R. Mazza[1,2], Qiangsheng Lu[1], Isaiah Gray[3], Qi Tian[3], Qinwen Deng[3], Duncan Moseley[1], An-Hsi Chen[1], Jason Lapano[1], Jason S. Gardner[1], Gyula Eres[1], T. Zac Ward[1], Erxi Feng[4], Huibo Cao[4], Valeria Lauter[4], Michael A. McGuire[1], Raphael Hermann[1], David Parker[1], Myung-Geun Han[5], Asghar Kayani[6], Gaurab Rimal[6], Liang Wu[3], Timothy R. Charlton[4], Robert G. Moore[1], Matthew Brahlek[1*]

[1]Materials Science and Technology Division, Oak Ridge National Laboratory, Oak Ridge, TN, 37831, USA
[2]Materials Science and Technology Division, Los Alamos National Laboratory, Los Alamos, New Mexico 87545, USA
[3]Department of Physics and Astronomy, University of Pennsylvania, Philadelphia, PA, 19104, USA
[4]Neutron Scattering Division, Oak Ridge National Laboratory, Oak Ridge, TN, 37831, USA
[5]Condensed Matter Physics and Materials Science Department, Brookhaven National Laboratory, Upton, New York 11973, USA
[6]Department of Physics, Western Michigan University, Kalamazoo, MI, 49008, USA
Correspondence should be addressed to *brahlekm@ornl.gov

**Abstract**: The field of spintronics has seen a surge of interest in altermagnetism due to novel predictions and many possible applications. MnTe is a leading altermagnetic candidate that is of significant interest across spintronics due to its layered antiferromagnetic structure, high Neel temperature ($T_N \approx 310$ K) and semiconducting properties. We present results on molecular beam epitaxy (MBE) grown MnTe/InP(111) films. Here, it is found that the electronic and magnetic properties are driven by the natural stoichiometry of MnTe. Electronic transport and *in situ* angle-resolved photoemission spectroscopy show the films are natively metallic with the Fermi level in the valence band and the band structure is in good agreement with first principles calculations for altermagnetic spin-splitting. Neutron diffraction confirms that the film is antiferromagnetic with planar anisotropy and polarized neutron reflectometry indicates weak ferromagnetism, which is linked to a slight Mn-richness that is intrinsic to the MBE grown samples. When combined with the anomalous Hall effect, this work shows that the electronic response is strongly affected by the ferromagnetic moment. Altogether, this highlights potential mechanisms for controlling altermagnetic ordering for diverse spintronic applications.

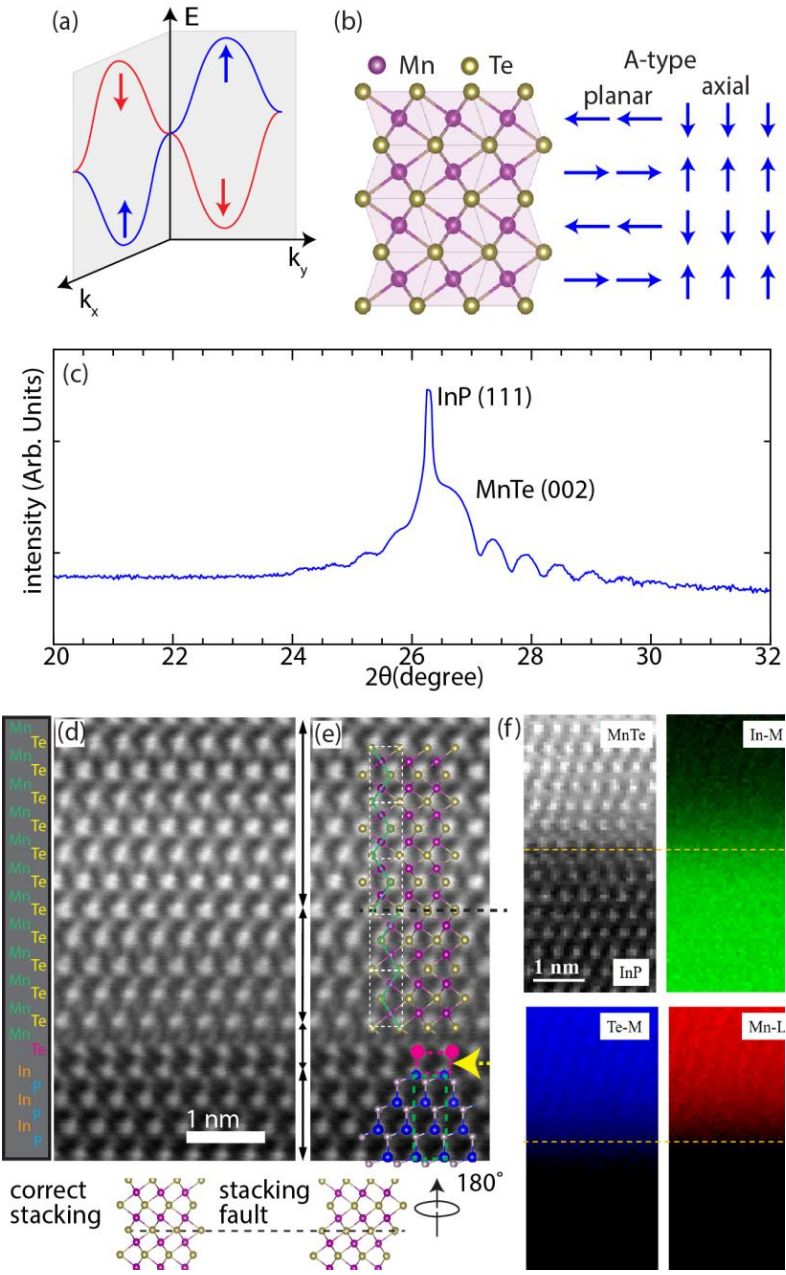

**Fig. 1** (a) Schematic of novel spin-momentum coupling predicted in a generic antiferromagnetic altermagnet. Here, this is depicted as a large spin splitting of the electronic band structure away from the high symmetry points. Adapted from Ref. [6]. (b) MnTe crystal structure (left) as well as the layered antiferromagnetic orderings with two known anisotropies. (c) X-ray diffraction $2\theta$-$\theta$ scans about the MnTe (002) reflection and the InP (111) reflection, as labeled. (d-e) HAADF-STEM image of InP/MnTe interface (see Fig. S1 for wide-angle view). Panel (e) shows the same image with overlaid structural models which highlight the atomic positions of the InP, the In-Te interfacial layer as well as the NiAs MnTe structure. (f) STEM-EELS spatial maps showing the atomic profiles across the interface.

Control of magnetism is essential for a wide range of technologies, from spin-based electronic devices[1] to emerging thermoelectrics[2]. Especially critical are magnetic thin film materials, as these are the key platform for most magnetic devices in current technologies. To push towards new applications requires understanding new magnetic phenomena and how to control both known and emerging material platforms as atomic-scale thin films. There has been rising interest in applications of antiferromagnets due to possibilities for fast switching speeds, lack of coupling to dipolar fields, and lower power operation [3,4]. Additionally, there are exciting new phenomena derived from antiferromagnetism that may enable bridging fundamental studies and applications. Of significant recent interest is the idea of *altermagnetism*, which is an electronic phase that is a consequence of the collinear antiferromagnetic order and the spin and crystalline symmetries[5–8]. The combination of rotational symmetries obeyed by the lattice and time-reversal symmetry allow for strong spin splitting of the band structure away from high-symmetry points of the Brillouin zone, as schematically shown in Fig. 1(a), which hold profound applications for spintronics such as spin splitting torque[9], spin hall effect[10], and tuning magnetoresistance[8]. If material routes can be found to understand and ultimately control altermagnetic materials, new antiferromagnetic-based technologies may become possible, which take advantage of the strong-spin momentum coupling without deleterious fringing fields found in ferromagnets.

MnTe is a prominent candidate altermagnet where a large spin-splitting has been predicted[6,11] and it is of significant interest for antiferromagnetic applications ranging from spintronics[3,12] to thermoelectrics[13,14]. Preliminary evidence of the band spin-splitting has also been reported on MnTe single crystals and thick films with angle-resolved photoemission spectroscopy (ARPES)[15–17]. Bulk MnTe crystallizes in the NiAs structure[18,19], which is hexagonal in-plane with Mn octahedrally coordinated by Te, as shown in Fig. 1(b) (see also Ref.[20] for review of early works on MnTe). Although bulk MnTe forms only in the NiAs structure, thin films have been stabilized in the cubic ZnS zinc-blende structure[21,22], as well as additional polymorphs. Electronically, MnTe in the NiAs structure (shown in Fig. 1(b)) is insulating with a direct band gap of ~1-1.5 eV[23], whereas the ZnS structure has a large gap of order 2-3 eV[22,24–26]. Magnetically, both the NiAs structure and the ZnS structure are antiferromagnetic[21,27]. In contrast to the ZnS structure, which exhibits a more complex ordering and low Neel temperature of $T_N \approx 70$ K[21,27], the NiAs structure of MnTe exhibits alternating ferromagnetic layers that are antiferromagnetically coupled along the c-axis with planar anisotropy[28]. The NiAs structure, as shown schematically in the right side of Fig. 1(b), moreover, exhibits a relatively large Neel temperature of $T_N \approx 307\text{-}310$ K[29,30]. One of the key aspects that make MnTe of critical interest to antiferromagnetic electronics is that the magnetic ground state is quite tunable. Due to small in-plane anisotropy, the in-plane spin-flop field is of the order of several Tesla[12,31]. Furthermore, controlling the magnetic anisotropy, and, thus, the magnetic space group, may be critical for accessing altermagnetic properties[6]. Here, a planar-to-axial magnetic anisotropy can be readily accessed via Li doping[32], which should have critical implications on the altermagnetic properties via changing spin textures, and further control the spin and electron transport properties. Moreover, a critical question is what happens to altermagnetic properties at high-quality surfaces and interfaces[33]. Surfaces are necessary for techniques like ARPES, which is the most straightforward way to confirm and understand the spin-momentum coupled electronic structure, though in situ cleaving bulk materials like MnTe may be a challenge as it is not inherently 2D. Moreover, surface and interfaces are present in all devices, and, thus, it is critical to understand novel device physics predicted for altermagnets. As such, the key question is how can the novel properties of MnTe be manipulated and better understood when synthesized as high-quality thin films.

Here, we show that molecular beam epitaxy (MBE) MnTe grown on lattice-matched InP possesses both a weak ferromagnetic moment and is metallic throughout the bulk of the film. This conclusion is reached by a combination of neutron diffraction, polarized neutron reflectivity (PNR), ARPES, and magnetotransport. The antiferromagnetic onset is close to the bulk-limit with $T_N \approx 310$ K, and the ferromagnetic moment onsets slightly lower than $T_N$. Temperature-dependent magnetotransport (magnetoresistivity and Hall effect) shows a strong anomalous Hall effect, which, taken with the other data suggest the ever-present weak ferromagnetic moment is a natural explanation for the origins of the anomalous Hall effect. Consistent with bulk crystals of MnTe and other members of the NiAs family, the magnetic response is directly linked to intrinsic doping effects due to excess Mn. This shows the natural magnetic state in MnTe and highlights how MnTe is tunable and can enable tailoring the ground state for diverse applications.

Figure 1(c) shows x-ray diffraction on an MBE-grown thin film (see supplemental section 1 for details of the growth) using a 4-circle X-ray diffractometer using copper $k_{\alpha 1}$ radiation. Here, the sharp peak is the (111) reflection of InP and the broader peak is the (002) reflection of NiAs MnTe. This confirms the NiAs phase, where any parasitic ZnS structure would be seen by the (111) reflection at around $2\theta \approx 24°$. The weaker fringes about the (002) reflection are Laue oscillations due to coherent diffraction off the top and bottom surfaces, indicating the crystalline interfaces are near atomically sharp. The high structural and interfacial quality enabled both neutron reflectometry and neutron diffraction to be performed to understand the magnetic and electronic properties. The depth-resolved structure of epitaxial MnTe on InP was directly probed using electron microscopy in Fig. 1(d-f). Figure 1(d) shows a high-angle annular dark-field scanning transmission electron microscopy (HAADF-STEM) image of the interfacial region, with wide-angle views shown in Fig. S1 (see supplement section 2). The same image is shown in Fig. 1€ but with the structural model overlain to not obscure the data in the image. Several things can be seen from this data due to the mass-difference of the elements. First, InP (bottom) can be clearly identified as the cubic a-b-c stacking sequence typical for (111) oriented ZnS structure. This is highlighted by the green-dashed box, as well as

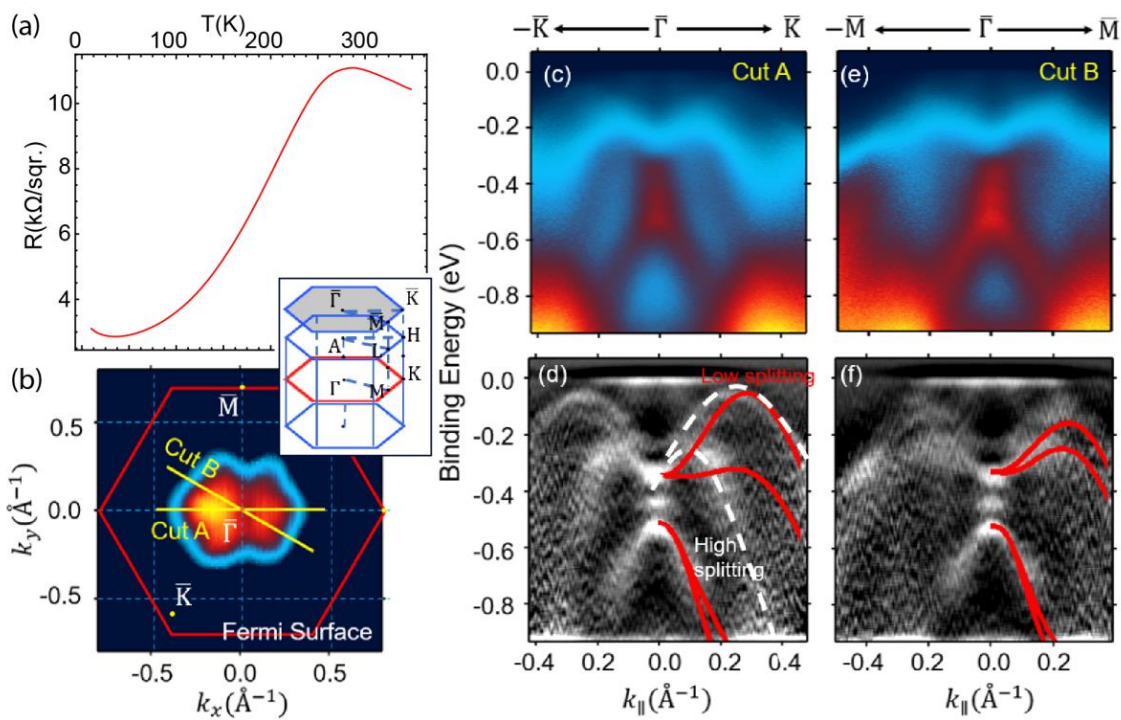

**Fig. 2** (a) Sheet resistance versus temperature for a 20 nm MnTe sample. (b) ARPES Fermi surface mapping, obtained using 11eV photon energy, the blue hexagon is the surface projected Brillouin zone (BZ) shown in the inset. (c-f) ARPES results along cut A (c,d) and cut B in (e,f), which is along ($\overline{\Gamma} - \overline{K}$) and ($\overline{\Gamma} - \overline{M}$) direction respectively. Panels (d,f) show corresponding curvature analysis plots from the ARPES spectra in panels (c, e) with overlays of the first principles calculated band structures along ($\overline{\Gamma} - \overline{K}$) and ($\overline{\Gamma} - \overline{M}$).

the dark atoms being the lighter P and the bright atoms being the heavier In, as labeled on the left-hand side. MnTe (top), in contrast, shows a hexagonal a-b stacking sequence typical of the NiAs structure, which is octahedrally coordinated. Moving upward and following the a-b sequence, a stacking fault can clearly be seen, as marked by the dashed black line. The overlain structure in panel(e) was found to fit only if the structure at the fault was rotated 180°, as highlighted at the bottom of the panel. Interestingly, for this local region, the structure appears to have an a-b-c stacking sequence, but it is clearly the octahedrally coordinated NiAs structure throughout, consistent with x-ray diffraction where there is a large peak shift for tetrahedrally coordinated ZnS structure.

The interface structure is nicely resolved and provides additional details regarding the overall properties of MnTe and how it grows on InP. First, the contrast of Mn, Te, In, P enabled resolving the sharpness of the interface. In moving upward in the image In-P-In-P ordering clearly terminates with In, which can be seen as a difference in the brightness of the atoms. The terminating In layer is clearly bonded vertically to Te. Since both Te and In are close in mass, this gives rise to the layer with near-vertical dumbbell shapes, highlighted by magenta balls and an arrow. This structure is reasonable given that the starting InP (111) termination was In (so-called (111)A) and the substrates were then exposed to Te for 15 minutes prior to growth. Furthermore, STEM electron energy loss spectroscopy (EELS) was used for element selective mapping across the interface, which is shown for In, Te, and Mn in Fig. 1(f). Here, all elements show relatively sharp truncation at the interfaces. However, it appears that there is a small amount of In within the MnTe layer, which could be intrinsic to the growth on InP (as is found in GaAs/MnAs[34,35]) or it is plausible that it could be an artifact of the focused ion beam sample preparation. In contrast, Mn and Te are quite sharp across the interface. These data agree well with x-ray diffraction and show that the films are of good structural quality with sharp interfaces.

As MnTe is a magnetic semiconductor, small amounts of doping strongly affect the electronic and magnetic properties. To show the presence of defects, the sheet resistance versus temperature is shown in Fig. 2(a) for a MnTe film that was 20 nm thick, and the measurement was performed in van der Pauw geometry with contacts made using pressed indium wire. Between 350 K and room temperature the resistance gradually increased with decreasing temperature, which indicates that the sample is weakly insulating. However, the rate of increase is much lower than would be expected for thermal activation since MnTe has a nominal bandgap of ~1.1 eV and the InP substrate was highly insulating (resistivity~$10^6$-$10^7$ $\Omega$cm[36]). Near 290 K, this trend is reversed, and the resistance reaches a broad local maximum, and starts to decrease with decreasing temperature, characteristic of a metal. This weak insulator-to-metal behavior is quite unusual and has been reported previously in thin films[12,37] as well as in doped bulk crystals[38]. It is, however, not a proper insulator-to-metal transition like those seen in prototypical systems like in vanadium-based oxides[39] or the rare earth nickelates[40], but in bulk crystals, it does coincide with the onset of antiferromagnetic ordering as well as a contraction along the c-axis[41]. For temperatures below ~250 K, the resistance continues to monotonically decrease with decreasing temperature until about 20 K, where there is a slight resistance upturn, consistent with disordered-driven localization.

ARPES measurements were performed on MnTe films transferred *in situ* from the MBE system (see supplement section 3 for details), and the data are shown in Fig. 2(b-f). As has been previously reported this anomalous metallic phase was proposed to be linked to the MnTe/InP interface[37]. However, as Fig. 2(b) shows there is a clear hexagonal Fermi surface, which was determined by integrating spectral intensity within a 20 meV window centered at the Fermi energy with sample $T$ = 8 K. Figure 2(c,e) shows energy-momentum cuts along the high symmetry lines in Fig. 2(b) and a curvature analysis plot of the same data (d,f), respectively[42]. Here, there are several bands below the Fermi level and the states at the Fermi level are p-type. This is consistent with bulk crystals which exhibit native p-type doping. It should be noted that the curvature analysis emphasizes bands below the Fermi level, but the bright, constant energy feature at the Fermi level in Fig. 2(d,f) is not a band but an artifact of the curvature analysis due to the Fermi-Dirac distribution[42]. The experimental and theoretical band structures agree well along the high symmetry lines Γ-K and Γ-M, where the red lines in Fig. 2(d, f) are theory (detailed results of the calculation are discussed in the supplementary information section 4). It is also found that the band-splitting along the Γ-K direction (marked by white dashed lines) is larger than the calculated band. This larger splitting is expected if the $h\nu$ = 11 eV photon energy is not centered at Γ (i.e. $k_z = 0$) but at a higher $k_z$. Because of the nature of the altermagnetism, the band spin-splitting is larger away from the high symmetry $k_z = 0$ plane, which is consistent with ARPES measurements on single crystal MnTe[15] and thick films[16].

A critical message of the data shown in Fig. 2, is that, despite the fact that it is expected to be an insulator, MnTe is metallic and exhibits a substantial hole-like Fermi surface. Relative band alignment and interfacial defects[43,44], or In-up diffusion might be possible dopants. Yet, consistent with thin films of MnTe grown on various substrates, this may point out that this property may be intrinsic to the Mn-Te system when grown by MBE (see for example[12,31,37,45,46]). The idea that the origin is related to intrinsic defects is consistent with similar NiAs compounds such as MnBi[47] or MnSb[48], where up to 13% excess Mn is readily accommodated interstitially. Moreover, the binary phase diagram for MnTe[20] suggests that MnTe should grow slightly Mn rich, implying that Mn-excess should be ever-present in crystals and films. This is also borne out in reports of bulk crystals where the properties can vary widely based on growth conditions[38,49,50]. As discussed later, this correlates well with an observed Mn-richness extending from the interface to the surface, as shown by PNR.

Given the known sensitivity of the magnetism in MnTe to strain and charge effects and the predictions of altermagnetism, the question arises—what is the nature of the magnetism in this system? To address this question, we performed neutron diffraction on a ~5×5 mm$^2$, 70 nm thick sample (see supplemental section 5). For MnTe the (001) peak is structurally forbidden but arises due to the A-type antiferromagnetic ordering with planar anisotropy. Figure 3(a) shows neutron diffraction scans of the (001) peak for temperatures of 50 K (blue circles) and 310 K (red triangles). The large difference in intensity among the magnetic but not structural peaks (see supplement Fig. S3) confirms the bulk of the MnTe films are in an antiferromagnetic phase with planar anisotropy. The temperature dependence of the intensity of the (001) peak is shown in Fig. 3(b). These data show a rise in intensity with reducing temperature that follows a general Bloch-type $T^N$ scaling. Moreover, this indicates that the Neel temperature of MnTe films is consistent with the bulk value of ~310 K.

To probe and depth resolve both chemical composition and in-plane ferromagnetism we performed neutron reflectivity on the MAGREF beamline at the Spallation Neutron Source[51] on a 20 nm film that was ~10×10 mm$^2$. Fig. 4(a) shows the reflectivity for neutrons with spin aligned with the in-plane magnetic field (4.5 T; $R^+$, blue squares) and those that oppose the field ($R^-$, black circles) along with the fits (solid lines, see supplementary section 5), and Fig. 4(b) shows the resulting spin-asymmetry $SA = (R^+ - R^-)/(R^+ + R^-)\times 100\%$. The overall shape and falloff of the reflectivity in Fig. 4(a) indicate that the films exhibit atomically sharp interfaces. In contrast to the x-ray reflectivity which show homogeneous oscillations (Fig. S5), the PNR data qualitatively exhibit a more complex chemical depth profile since neutron reflectivity is very sensitive to the chemical and isotopic species. For MnTe, the relevant individual coherent scattering lengths are quite distinct, e.g., Mn: -3.73 fm, Te: 5.80 fm, O: 5.80 fm, In: 4.07 fm and P: 5.13 fm. As detailed in the supplement section 5, this can be made quantitative by modeling the $R^+$ and $R^-$ reflectivity. These models were co-refined using Bayesian analysis using Refl1D, which gives the depth resolved nuclear scattering length density (NSLD), and the magnetic scattering length density (MSLD). These results were obtained using a three-layer model that consisted of an interface region, the MnTe film with a slight chemical gradient and an oxide layer on the surface, as shown in the schematic at the top. The solid lines in Fig. 4(c) are the profiles, the shaded areas about the lines are the confidence intervals.

The fit shows that the films are likely Mn-rich with a gradient progressing towards the surface, as indicated by the net negative NSLD. This deviation from perfect 1-to-1 stoichiometry is directly confirmed using Rutherford backscattering spectroscopy (RBS) measurements, which show that the films are naturally ~10% Mn-rich (See Fig. S6). Similarly, the high SLD near the surface indicates the presence of an oxide layer, as shown by the large upturn in NSLD due to the density enhancement by the addition of oxygen[52]. Here, the surface oxide layer is ~6-7 nm thick with a roughness of ~2-3 nm at the MnTe interface. This suggests that there may be a kinetic limitation for the oxygen diffusion at room temperature, which considering the strong effect on materials with similar A-type antiferromagnetic ordering[52], is a very interesting point for future study. On the surface there is an additional layer where there is a low scattering length density. Since this is significantly larger that thickness extracted from x-ray reflectivity, this is most likely a thin layer accumulated within the cryostat, as it is absent in room temperature PNR data.

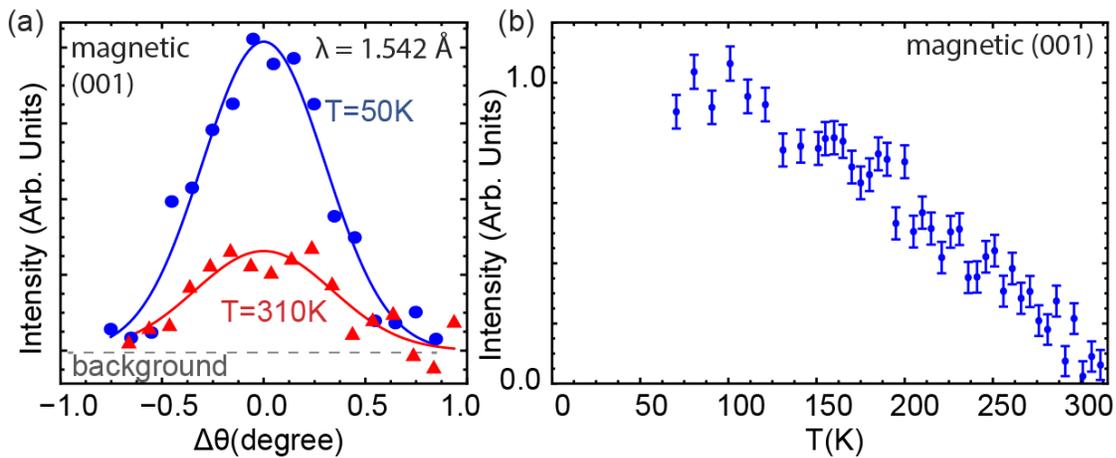

**Fig. 3.** (a) Neutron diffraction rocking curves of the structurally forbidden (001) peak at 310 K (red) and 50 K (blue). The solid curves are guides to the eye. (b) Intensity of the magnetic (001) peak versus temperature.

Magnetically, the spin-asymmetry is clearly non-zero at 1.5 K, which shows definitively that there is a net ferromagnetic moment in the films. This contrasts with the zero SA at 300 K, indicating the net moment is gone (supplementary Fig. S4). The model for the MSLD suggests that the magnetism exists throughout the bulk of the MnTe layer. There are two additional regions within the bulk MnTe layer where the layer closer to the InP interface has a larger moment, which likely follows the Mn-rich gradient extending away from the interface. The model further suggests that magnetism also exists within the surface oxide layer, but the uncertainty is quite large. The SA can be converted to net moment, which yields a value of roughly 12 emu/cm$^3$ for the film and 50 emu/cm$^3$ for the oxide layer, which roughly matches magnetometry measurements shown in Fig. S7, as well as a rough expectation for the magnetization resulting from percent-level excess Mn, assuming several $\mu_B$ per excess Mn atom. As MnTe possess intrinsic antiferrmagnetism and weak ferromagnetism, the question then arises if there is coupling among these states. The magnetometry measurements show clear exchange biasing. The character of the vertical shift of the loops shows that the ferromagnetic moments are pinned and thus dominated by the antiferromagnetism, which is consistent with weak ferromagnetism embedded in an antiferromagnet[53,54]. Altogether, the neutron diffraction data showed a bulk antiferromagnetic phase while reflectometry showed there exists a net ferromagnetic moment, which is likely linked to the metallic transport through the excess Mn naturally present in the films.

To understand the interrelation of the magnetism and the electronic properties, magnetotransport (Hall and resistance) measurements were made with the magnetic field oriented out of the plane of the sample ($H$ parallel to the <001> direction) and the data was symmetrized to eliminate inadvertent mixing of the longitudinal and Hall resistance. Figure 4(d) shows the anomalous Hall resistance, $R_{xy,A}$ (upper panel) and magnetoresistance $MR = (R(H)-R(H=0T))/R(H=0T)\times100\%$ (lower panel) for $T$ = 50 K and 200 K, which have been offset for clarity. The anomalous Hall effect was obtained by subtracting a line with slope $R_{xy,0}$ from the Hall effect, ie, $R_{xy,A}=R_{xy}-R_{xy,0}\times H$, where $H$ is the applied magnetic field. Ferromagnetic hysteresis is observed in both the $R_{xy,A}$ and MR and is found to onset between 250-275K (see Fig. S8 for data from full temperature dependence from 300 K down to 2 K), consistent with reports in bulk crystals. The data taken at 200 K shows a single loop for the anomalous Hall effect and a butterfly shaped curve found in the MR. At 50 K, the ferromagnetic hysteresis evolves to a double-loop in both the magnetoresistance and the anomalous Hall effect, indicating that the ferromagnetic component likely evolves at low temperature, perhaps following the gradient observed in PNR, interaction with antiferromagnetism as indicated by magnetometry measurements (Fig. S7), or coupling to the magnetism within the oxide layer. However, concerning the oxide layer, similar transport character is seen across all samples and does not change with capping layer or growth conditions. For example, Fig. S9 shows magnetotransport for an uncapped sample, whereas Fig. S8 is an uncapped sample. The similar behavior for this capped sample indicates this is an intrinsic response of the MnTe system and not due to any magnetism within the oxide layer as suggested by PNR.

As the altermagnetic candidate MnTe is a band insulator and the fact that it ubiquitously shows metallic transport points to existence of defects or extrinsic impurities. This likely explains many of the observed electronic and magnetic properties, and, thus providing new insight into how to synthesize materials with close-to-predicted properties as well as how to tailor materials to achieve a desired property. As we have shown, MBE grown films are Mn-rich, which, in fact, coincides with the thermodynamic expectation. This insight was obtained with RBS, and PNR profile suggests the films are Mn-rich with a slight gradient extending from the interface towards the surface. This agrees with transport and ARPES results which show that the bulk is metallic and p-type, consistent with previously reported carrier densities in the range of $10^{19}$ cm$^{-3}$ and overall low mobilities of order 10-100 cm$^2$V$^{-1}$s$^{-1}$. Regarding the magnetism, we find that films are antiferromagnetic with weak ferromagnetic moments. MnTe is intrinsically magnetic, and, thus, any native defects should have a strong effect on the magnetic properties; this is demonstrated in magnetometry measurements that imply a strong coupling among the ferromagnetic and antiferromagnetic states. Therefore, as schematically shown in Fig. 4(e), the defects likely point to either a small canting of the Mn atoms (i and ii) or related to local moments that arise around defect sites (iii). Yet, it remains an open question regarding the dominant interaction that determines the ferromagnetic coupling. The metallic nature and the weak ferromagnetism together imply a strong coupling of the electronic and magnetic responses. This is observed as a strong anomalous Hall effect, which is found to arise below the Neel temperature. With further reducing temperature, the character of the anomalous Hall is found to change, which is likely related to the Mn gradient observed in PNR or coupling to the bulk antiferromagnetism.

To conclude, this work highlights that MnTe thin films are highly tunable both electronically and magnetically, which are both tied to the fundamental character of the MBE growth of MnTe. These properties, when coupled with the predictions of altermagnetism, highlight many realistic avenues for designing new functional electronic/magnetic properties. Important questions remain. Specifically, what are the intrinsic defects that form in MnTe films (interstitials, vacancies, or antisite defects) and how such defects contrast with the range of magnetic and doping properties reported in bulk crystals, as well as films on many different substrates. Moreover, what role do surfaces and interfaces play in the novel properties of altermagnets[33] and how can these be leveraged to achieve a desired functional response. For example, is it possible to utilize the layered A-type magnetism to generate a net moment through even-odd thickness control, akin to recent work on the magnetic topological material $MnBi_2Te_4$. Moreover, is it possible to synthesize MnTe films that are both electrically insulating as well as lacking any ferromagnetic moment. In addition, how does the combination of charge doping, and ferromagnetic moment affect the proposed altermagnetic spin texture in the electronic bands, and what are appropriate growth conditions or doping schemes to achieve this level of control. This is critical to search for novel effects of altermagnetism, in particular, the anomalous Hall effect without a net moment[55]. Moreover, the results here highlight that any spontaneous anomalous Hall effect derived from an antiferromagnet is exceptionally hard to confirm and would require near defect-free materials. Pushing defect density down is challenging in candidate systems, given the high propensity of defects that tend to occur in highly covalent materials such as MnTe[56,57]. By addressing these open questions, many new functional responses will become available for study and pave the way for new magnetic devices and technologies.

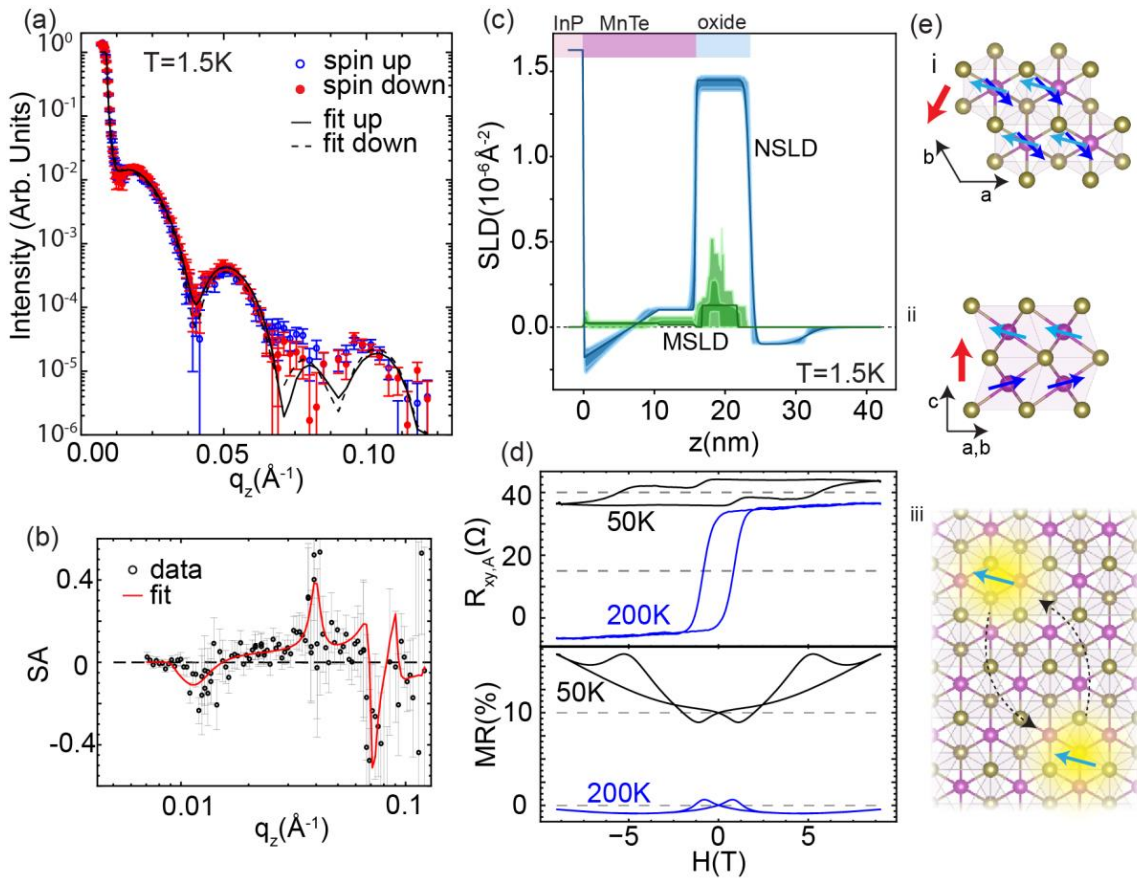

**Fig. 4.** (a-b) Polarized neutron reflectivity (a) curves for spin-up and spin-down (symbols) along with the spin asymmetry (SA, b) and corresponding fits (solid lines). (c) Depth resolved nuclear scattering length density (SLD) and magnetic SLD versus the thickness. These were extracted from the fits to the data in (a-b) at 1.5 K. (d) Magnetotransport (anomalous Hall effect, $R_{xy,A}$ top panel and magnetoresistance, *MR*, bottom panel) versus magnetic field at 200 K and 50 K; dashed lines indicate offsets. (e) Schematic of possible mechanism for ferromagnetism consistent with the experimental data. (i-ii) Slight canting of the net-antiferromagnetic ground state, or (iii) localized net moments with ordering mediated by long-range ferromagnetic interactions represented by the dashed black arrows.

## Data Availability

The data supporting this study's findings are available from the corresponding author upon reasonable request.

## Supplementary Information

Supplementary Information includes 1. Molecular beam epitaxy growth; 2. Electron microscopy ; 3. Angle-resolved photoemission spectroscopy; 4. First principles calculations; 5. Neutron scattering; 6. X-ray scattering.

## Acknowledgments

This work was supported by the U. S. Department of Energy (DOE), Office of Science, Basic Energy Sciences (BES), Materials Sciences and Engineering Division (growth, neutron scattering, transport, density functional calculations, analysis, and manuscript preparation). Portions of this work were supported by the U.S. Department of Energy, Office of Science, National Quantum Information Science Research Centers, Quantum Science Center (Q. L. and R. G. M., spectroscopy, analysis, and manuscript preparation). EF and HC acknowledge the support of US DOE BES Early Career Award No. KC0402020 for neutron diffraction and manuscript preparation. This research used resources at the High Flux Isotope Reactor and the Spallation Neutron Source, the DOE Office of Science User Facility operated by ORNL. The electron microscopy work at Brookhaven National Laboratory was supported by the U. S. DOE, Office of Science, BES, Materials Science and Engineering Division under Contract No. DESC0012704. Some of the neutron reflectivity was also supported by the NNSA's Laboratory Directed Research and Development Program at Los Alamos National Laboratory. Los Alamos National Laboratory, an affirmative action equal opportunity employer, is managed by Triad National Security, LLC for the U.S. Department of Energy's NNSA, under contract 89233218CNA000001. IG was sponsored by the Army Research Office under the grant No. W911NF-20-2-0166; QD and QT were supported by the National Science Foundation Electronic and Photonic Materials Program program under grant no. DMR-2213891 for optical measurements.

# Supplementary Information
# Stoichiometry-induced ferromagnetism in altermagnetic candidate MnTe

1. **Molecular beam epitaxy growth**

Thin films of MnTe were grown via molecular beam epitaxy (MBE) on (111)A InP substrate, where Fe-doping is standard to make the substrates electrically insulating (resistivity~$10^6$-$10^{-7}$ $\Omega$cm[36]). The substrates were terminated using a two-step method. This method consisted of using ozone to create an oxide layer and remove organics, which was followed by an etch in 1:10 $H_2O$:HCl to fully remove the oxide layer. The substrates were then quickly placed in deionized water. They were then removed, blown dry, mounted on the sample plate with silver paste, then placed under flowing $N_2$ gas and heated to cure the Ag paste. The samples were then pumped down and transferred into the MBE system. This was undertaken with air exposure kept to less than about 30 seconds. The substrates were then heated to the growth temperature. Te was opened at 250°C and remained open throughout the growth. At these elevated temperatures any excess Te desorbed off the surface. At 250°C a 2 monolayer seed layer was initially grown. Following this, the films were heated to 325°C where the remainder of the film was grown to the desired thickness. This growth process resulted in repeatable growths with good structural quality and was critical to achieve the atomically flat surfaces necessary for PNR.

2. **Electron microscopy**

High-angle annular dark-field scanning transmission electron microscopy (HAADF-STEM) specimens were prepared using focused ion beam (FIB) and lift out with a FEI Helios G5 UX DualBeam FIB/scanning electron microscope with final Ga+ milling performed at 2 keV. HAADF-STEM and electron energy loss spectroscopy (EELS) were performed at Brookhaven National Laboratory, using a JEOL ARM 200CF equipped with a cold field emission gun and spherical aberration correctors operated at 200 kV.

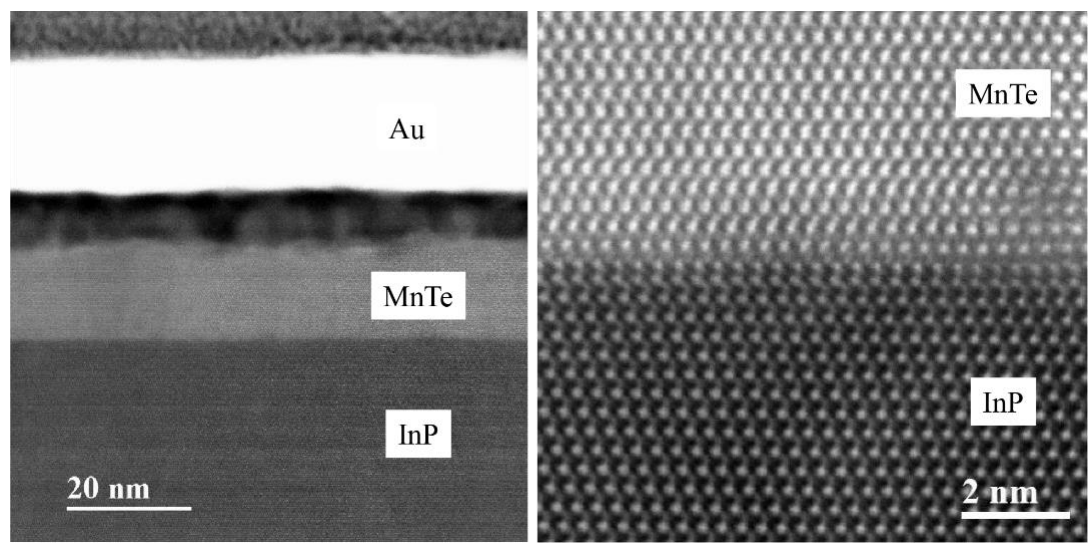

**Fig. S1**. Wide-scale HAADF-STEM images of InP/MnTe interface. The dark layer between the Au and MnTe is damage during the electron microscopy sample preparation as confirmed by the thickness measured using X-ray reflectivity of the same sample.

3. Angle-resolved photoemission spectroscopy

After the growth, the MnTe films were transferred *in situ* to the ARPES system maintained at a base pressure of $< 5\times10^{-11}$ mbar. ARPES measurements were performed at $T \sim 8$ K using a Scienta DA30L hemisphere analyzer with the sample illuminated with an Oxide $hv = 11 eV$ laser system, which gives an estimated probing depth of roughly 1 nm, similar to soft x-ray ARPES reported in [16]. The light in this experiment is linearly polarized and perpendicular to the sample and the slits of the detector (i.e. p-polarization). The system was configured with a pass energy of 5 eV and 0.3 mm slit yielding an energy resolution ~4 meV, and the momentum resolution ~0.01 Å$^{-1}$. The size of the beam spot on the sample was ~100 μm and we did not observe any changes in the ARPES spectra when changing the beam position on the sample surface, indicative of the homogeneity of the grown samples.

4. First principles calculations

First-principles calculations with density functional theory (DFT) were performed using the Vienna ab Initio Simulation Package (VASP) package[59]. We used the Perdew–Burke–Ernzerhof (PBE)[60] form of the exchange-correlation functional. All the calculations were performed with a plane-wave cut-off energy of 300 eV on the 10×10×10 Monkhorst-Pack k-point mesh. MnTe with an in-plane lattice constant of 4.15 Å and an out-of-plane lattice constant of 6.71 Å. The atomic positions were optimized by the conjugate gradient method. Calculations of the band structures were performed with the inclusion of spin-orbital coupling (SOC). Different collinear ferromagnetic and antiferromagnetic (AFM) orders were calculated, and the lowest energy state found was the A-type in-plane antiferromagnetic order along (110) direction, which is consistent with experimental reports. The band structures overlayed in Figure 2 (main text) are calculated along **Γ**-K and **Γ**-M direction, including SOC and (110) A-type AFM order.

First principles calculations of the electronic structure of bulk MnTe, within the generalized gradient approximation of Perdew, Burke and Ernzerhof [4], along with a U of 5 eV applied to the Mn 3$d$ orbitals, were also conducted within an all-electron approach, using the linearized augmented planewave density functional theory code WIEN2K[61]. The band structure for these calculations is shown in Fig. S2 below. For these calculations, 1000 k-points in the full Brillouin zone were employed, within the experimentally determined magnetic order (alternating planes of Mn atoms). An RK$_{max}$ of 9.0 was employed – the product of the smallest muffin-tin radius and the largest plane-wave expansion wavevector, with muffin-tin radii of 2.50 Bohr were employed for Mn and Te, with lattice parameters of 7.818 Bohr (plane) and 12.693 Bohr (c-axis). Spin-orbit coupling was not included.

We note that the inclusion of a U is key to obtaining a properly insulating state as with the regular GGA metallic behavior, contradicting the well-known insulating character. One notable feature of the electronic structure is that along many high-symmetry directions the spin splitting characteristic of an altermagnet vanishes (see Fig.S2), but band splittings as large as several tenths of an eV are obtained along other low-symmetry directions, such as Γ to L or M to A. This is likely a consequence of the highly symmetric structure of MnTe (the unit cell has 24 symmetry operations). We defer more detailed discussion of this behavior to a subsequent paper.

This insulating, antiferromagnetic behavior is at odds with the metallic ferromagnetic behavior we observe experimentally and speaks to the importance of the interface in the physics here; calculated ferromagnetic bulk Mn spin orientations, in general, fall hundreds of meV per Mn above this ground state.

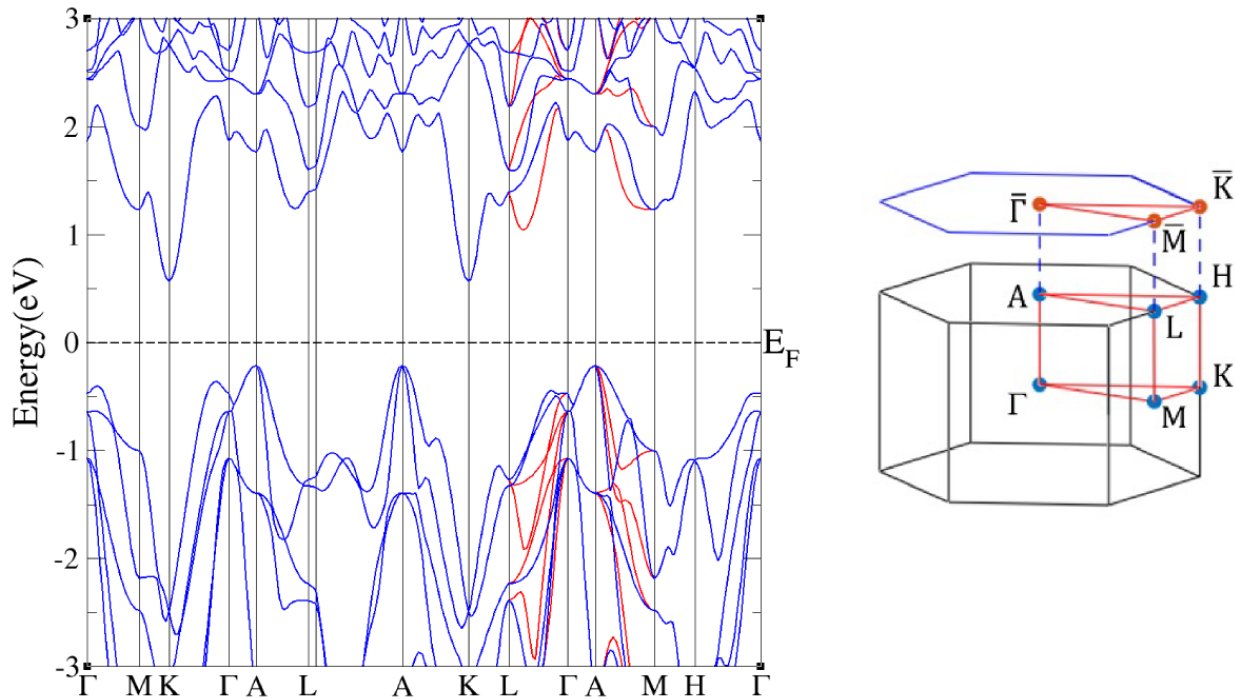

**Fig. S2**. Band structure and corresponding Brillouin zone calculated for MnTe with a $U = 5$ eV, consistent with the experimental bandgap of ~1eV.

## 5. Neutron scattering

Neutron diffraction was done at the HB-3A DEMAND using a wavelength of 1.542 Å[58] at the High Flux Isotope Reactor (HFIR) at Oak Ridge National Laboratory (ORNL).

Polarized neutron reflectometry was performed at the Spallation Neutron Source (SNS) at ORNL on the magnetic reflectometry (MAGREF) at beamline 4A. The $R^+$ and $R^-$ reflectivity were corefined using the Refl1D program. The use of Refl1D allows for Bayesian analysis of the posterior distributing produced in the fitting process. Confidence intervals are extracted for each refinable parameter. This enables understanding of the significance of features such as those in the magnetic scattering length density (SLD) in Fig. 4 and Fig. S4. As shown, the statistical likelihood of observing a non-zero moment is negligible. This enhances our confidence in claiming zero net moment at room temperature as compared to temperatures below $T_N/T_c$.

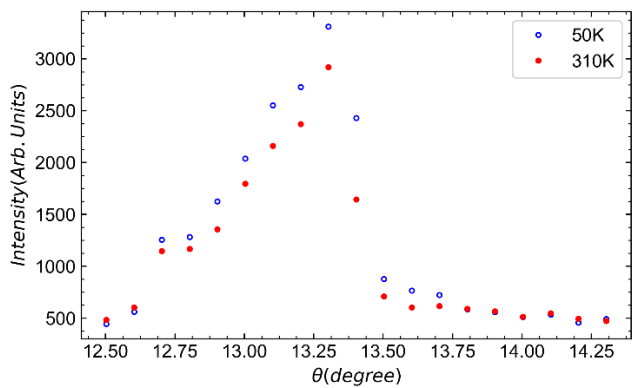

**Fig. S3**. Neutron diffraction rocking curves of the combined structural and magnetic 101 peak at 310 K (red) and 50 K (blue). This peak shows minimum change of intensity with reducing temperature which contrasts the pure magnetic peak shown in Fig. 3 of the main text.

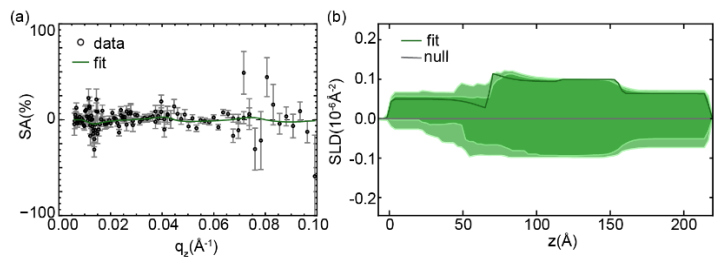

**Fig. S4**. (a) Spin asymmetry measured at room temperature. (b) Magnetic scattering length density extracted from the fitting in (a) versus the thickness, along with a null result (no ferromagnetism).

## 6. X-ray scattering

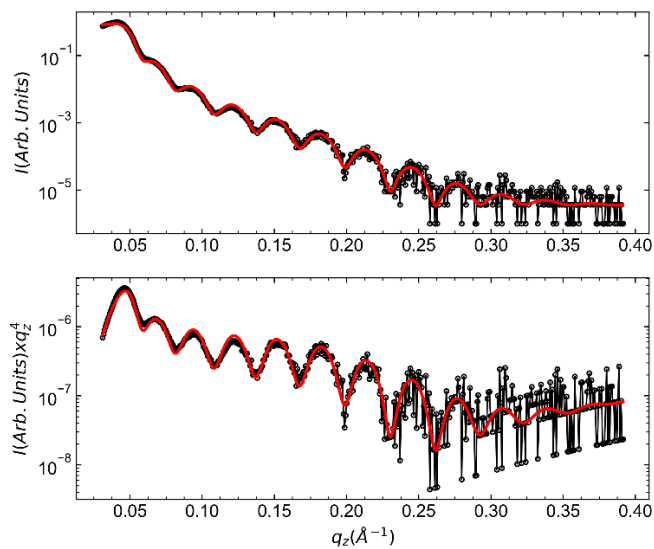

**Fig. S5** X-ray reflectivity and single-layer model versus $q_z$ for the same sample shown in the main text. The upper plot is the normalized intensity and lower plot is the intensity multiplied by the standard $q_z^4$ for Fresnel reflectivity to highlight the level of homogeneity seen by x-ray scattering.

## 7. Rutherford backscattering spectroscopy

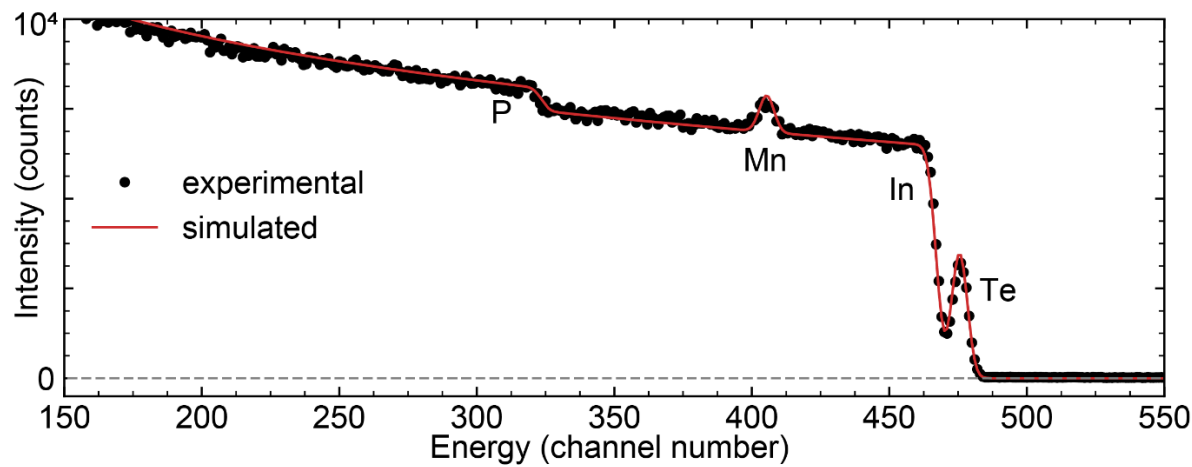

**Fig. S6** Rutherford Backscattering spectroscopy data (symbols) on a MnTe film grown on InP and the corresponding simulation (solid). Despite an overpressure of Te (>5:1) during the growth, the resulting average composition contained $4.5 \times 10^{16}$ cm$^{-2}$ Mn atoms and $3.8 \times 10^{16}$ cm$^{-2}$ Te atoms, which is in good agreement with the nuclear SLD composition from neutron reflectivity as well as the targeted thickness of 20 nm.

## 8. Magnetometry and exchange bias

To understand whether the moments contributing to a ferromagnetic (FM) response are directly coupled to those contributing to the bulk antiferromagnetism (AFM), we performed field-biased measurements. Here the sample was cooled in a large magnetic field and an M vs H field sweep was performed. This enabled understanding the interfacial coupling of two magnetic phases. When a sample has multiple magnetic phases, a simplistic view of field biased measurements can boil down to changes in magnetization vs field responses of four types: (1) Vertical shift resulting from the pinning of FM/uncompensated moments at an interface with a AFM (AFM dominated response), (2) horizontal shift resulting from the sluggish reversal of FM moments at an interface with AFM (FM, AFM competing on similar energy scales), (3) increase in the overall coercivity resulting from FM moments reversing the AFM moments at an interface (FM dominated), and (4) no change due to lack of direct coupling[53,54]. In our samples, we performed the measurements by field cooling at $H = +/- 7$ T from above the Neel temperature (350 K) to 4 K. As shown in Fig. S7, the resulting response is dominated by type (1) outlined above, a vertical shift in the magnetization. This indicates an imbedded FM matrix in a dominate long-range magnetic system, as has been seen in other magnetically frustrated systems[62]. This measurement not only confirms the presence of FM in the sample but also furthers our understanding of how it is dominated energetically by the percolated AFM phase.

Magnetometry experiments were conducted on a 50 nm film of approximate area of 0.04 cm$^2$ using a Quantum Design MPMS3. Substrate background was subtracted using a linear diamagnetic signal. Field cooled measurements were done by first warming the sample to 350 K followed by cooling in a +/- 7 T field to 4 K. Magnetization vs field measurements were then done in the out-of-plane configuration and completed by sweeping starting at the bias field (+/- 7 T).

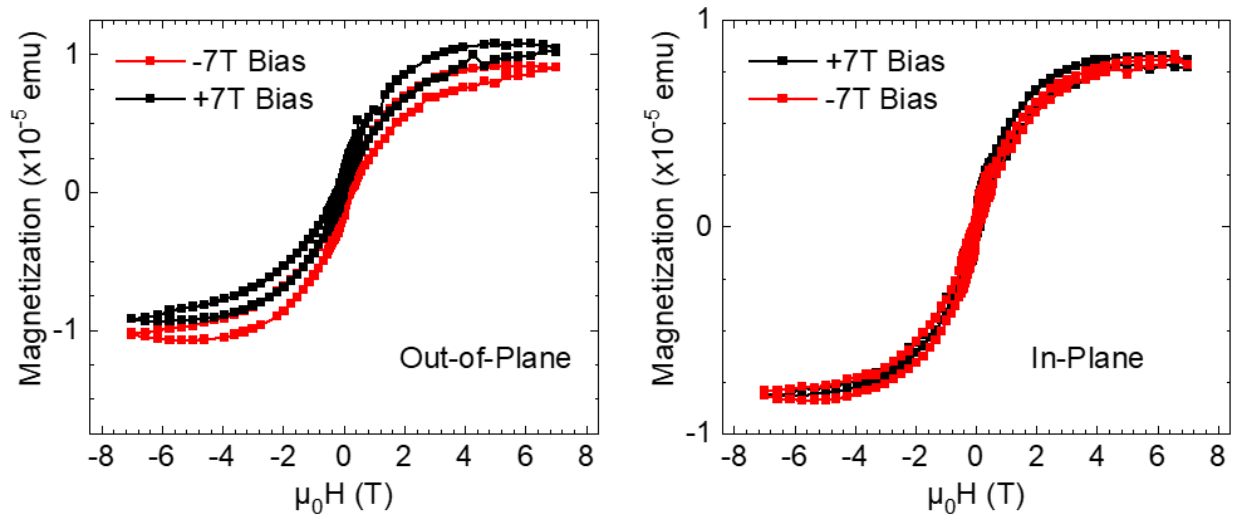

**Fig. S7** Magnetometry measurements with the magnetic field oriented out-of-plane (left) and in-plane (right). When field cooled at ±7 T, a small vertical shift is observed indicating exchange bias. The net saturation moment agrees well with the PNR results as well and the density of excess Mn measured using RBS.

## 9. Extended Transport data

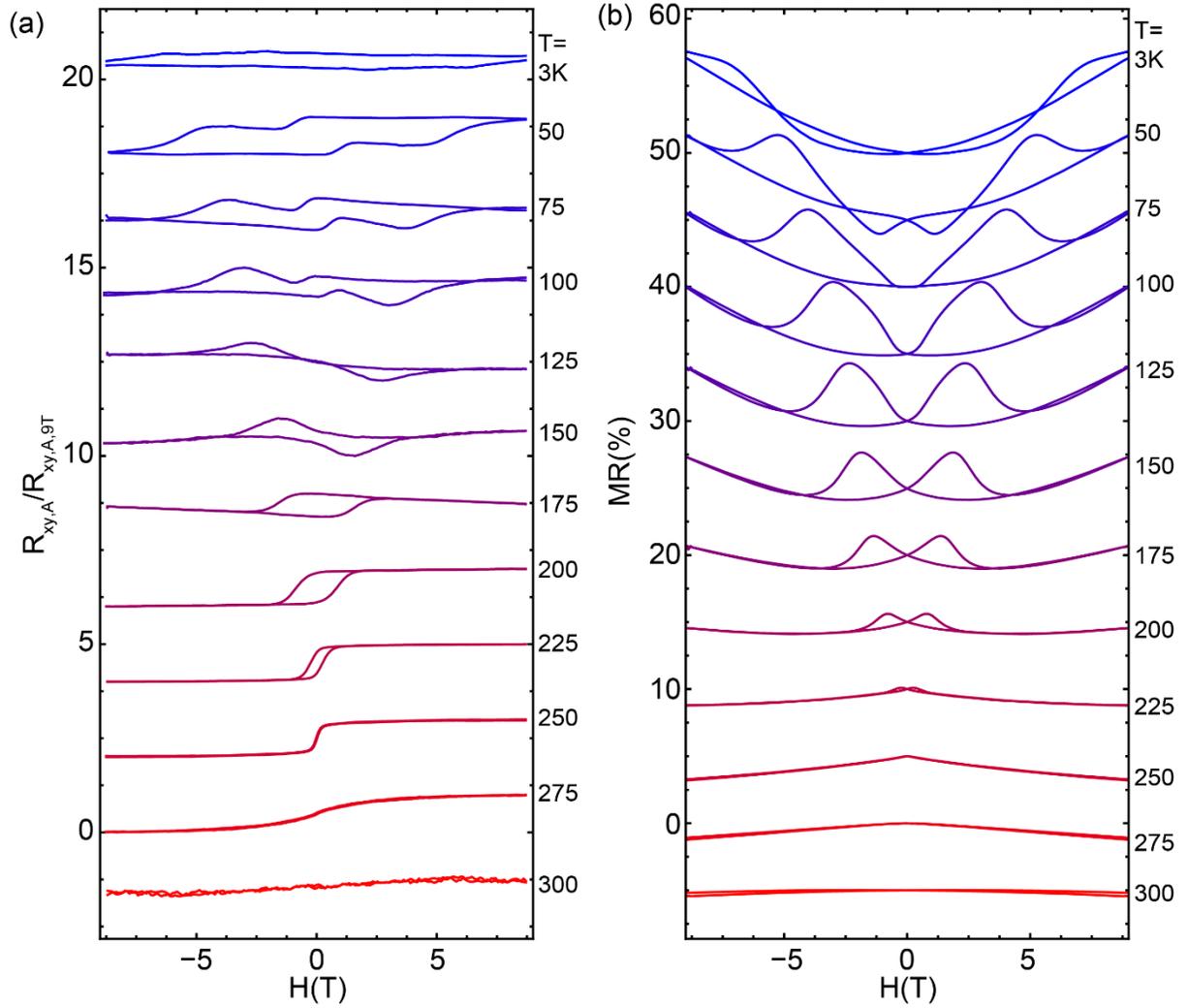

**Fig. S8** (a-b) Full temperature dependence of the magnetotransport (anomalous Hall effect normalized at 9T, (a) and magnetoresistance, MR, (b)) versus magnetic field. Here hysteresis develops between 275 K and 250 K, indicating that the Curie temperature is below the Neel temperature.

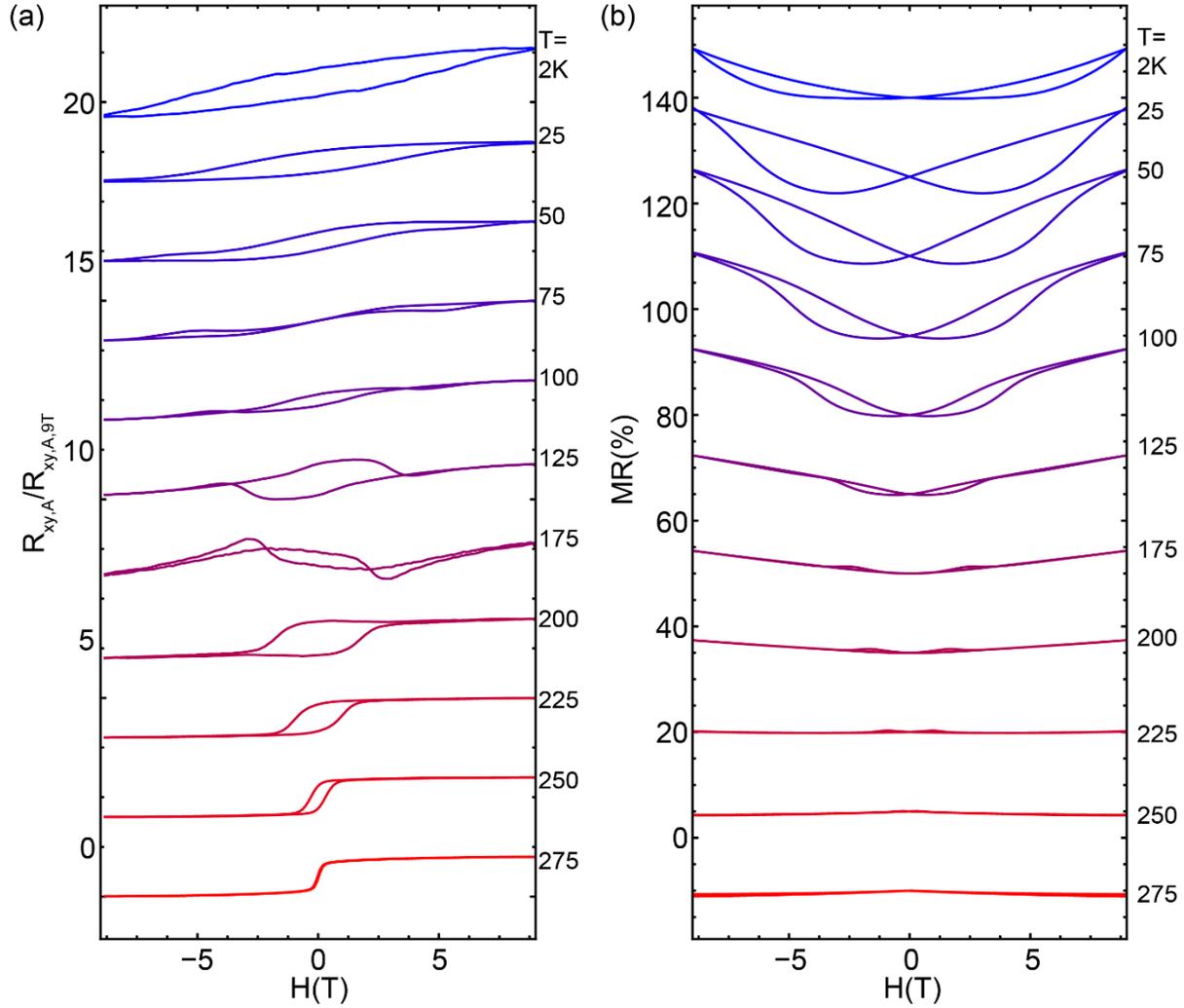

**Fig. S9** (a-b) Full temperature dependence of the magnetotransport (anomalous Hall effect normalized at 9T, (a) and magnetoresistance, MR, (b)) versus magnetic field. This data was with a Te capping layer (with no oxide), whereas the data in Fig. S8 was without a capping layer.

## 10. Circular Dichroism

To probe the combined magnetic and electrical response, CD was performed on a 40 nm Te capped sample, as reported in Ref.[63]. CD is proportional to the difference of index of refraction of left-hand and right-hand circularly polarized light, $\Delta N = i\varepsilon_{xy}/\varepsilon_{xx}$, where $\varepsilon_{xy}$ and $\varepsilon_{xx}$ are the transverse and longitudinal dielectric constant at the incident photon energy[64–66]. The dielectric and optical conductivity tensors can be related via $\varepsilon(\omega) = 1 + 4\pi i \omega \sigma(\omega)$. This relation can then be used to show that CD is proportional to $\sigma_{xy}$ at the optical frequency, and, thus, is an optical method to detect the anomalous Hall conductivity[67]. CD hysteresis loops were, therefore, performed with an out-of-plane applied field between ±5 T to measure potential surface magneto-optical effects arising from altermagnetism, surface ferromagnetic moments or both. The results are shown in Fig. S10 for a 40 nm thick film at four temperatures, using 800 nm laser wavelength where the curves have been offset by $2\times10^{-4}$ for clarity. The key finding is that no hysteresis was observed within measurement sensitivity (minimum CD $\approx 2\times10^{-5}$). Similar null results were obtained for a 20 nm thick sample at 633 nm wavelength. Therefore, no indication of altermagnetism or surface ferromagnetism was found. These null results suggest that the ferromagnetism observed in neutron reflectivity and in $\rho_{xy}(H)$ loops in Fig. 4 is very dilute and may reside below the magneto-optical sensitivity depth of the MnTe/Te system (although hard to determine, typically of the order of nanometers for degenerate doped semiconductors, which may be also reduced by the 20 nm Te capping layer). This is overall consistent with there being either a magnetic gradient extending away from the interface as proposed by neutrons or the net magnetic moment being extremely small, or a combination.

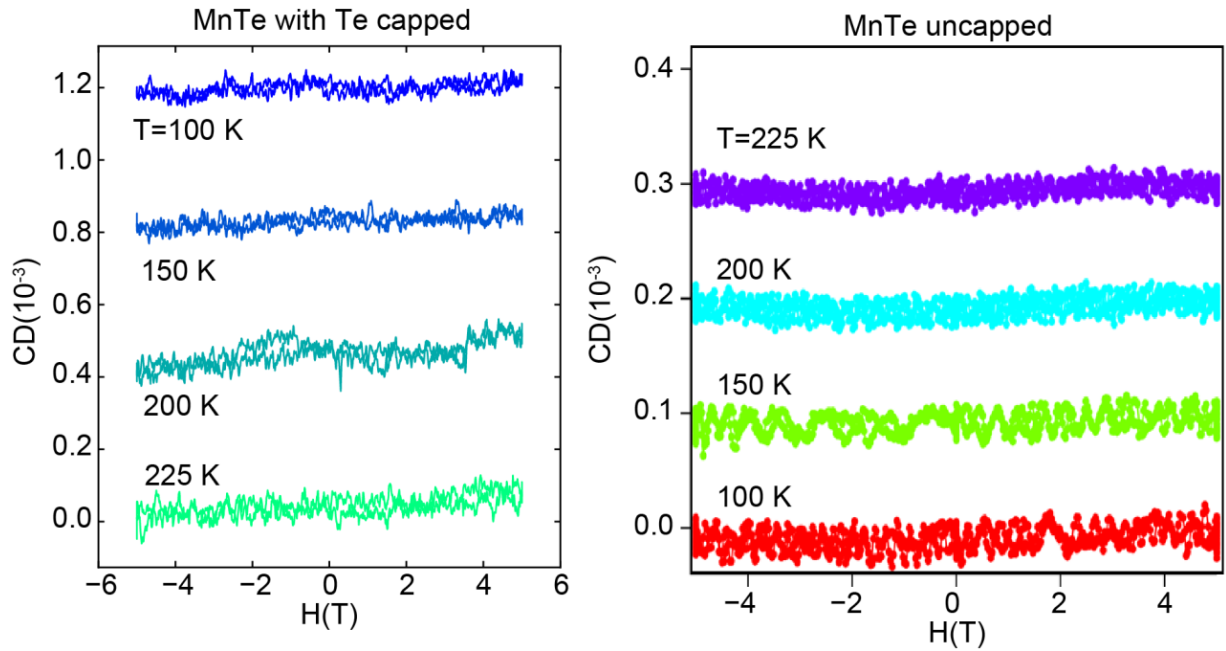

**Fig. S10** Circular dichroism hysteresis loops of a Te capped MnTe film (left) and an uncapped MnTe film (right) which were measured at 800 nm wavelength, with magnetic field applied out-of-plane.